\documentclass[letterpaper, 10 pt, conference]{ieeeconf}  

\IEEEoverridecommandlockouts                              

\usepackage{graphicx}
\usepackage{amsmath}
\usepackage{amssymb}
\usepackage{multicol}
\usepackage{cuted}
\usepackage{multirow}
\usepackage{caption, subcaption}

\newtheorem{problem}{Problem}
\usepackage{xcolor}
\usepackage{cite}
\usepackage[colorlinks=true,allcolors=steelblue]{hyperref}
\definecolor{steelblue}{RGB}{70,130,180}

\def\bbn{\mathbb N}
\def\bbz{\mathbb Z}
\def\bbr{\mathbb R}

\def\bbs{\mathbb S}
\def\bbp{\mathbb P}
\def\bbe{\mathbb E}

\newcommand{\Let}{: =}

\DeclareMathOperator{\vect}{vec}

\title{\LARGE \bf
Anomaly Detection Under Multiplicative Noise Model Uncertainty
}

\author{Venkatraman~Renganathan*, Benjamin~J.~Gravell*, Justin~Ruths, and~Tyler~H.~Summers
\thanks{*Equal contribution of these authors. This material is based on work supported by the United States Air Force Office of Scientific Research under award number FA2386-19-1-4073 and by the National Science Foundation under award number ECCS-2047040. V. Renganathan is with the Department of Automatic Control, Lund University, Sweden. B.J. Gravell, J. Ruths, and T.H. Summers are with the Department of Mechanical Engineering at The University of Texas at Dallas, Richardson, TX, USA. E-mail: {\tt\small venkat@control.lth.se, (benjamin.gravell, jruths, tyler.summers)@utdallas.edu.}}}

\begin{document}

\maketitle
\thispagestyle{empty}
\pagestyle{empty}

\begin{abstract}

State estimators are crucial components of anomaly detectors that are used to monitor cyber-physical systems. 
Many frequently-used state estimators are susceptible to model risk as they rely critically on the availability of an accurate state-space model. 
Modeling errors make it more difficult to distinguish whether deviations from expected behavior are due to anomalies or simply a lack of knowledge about the system dynamics. 
In this research, we account for model uncertainty through a multiplicative noise framework. 
Specifically, we propose to use the multiplicative noise LQG based compensator in this setting to hedge against the model uncertainty risk. 
The size of the residual from the estimator can then be compared against a threshold to detect anomalies. Finally, the proposed detector is validated using numerical simulations. 
Extension of state-of-the-art anomaly detection in cyber-physical systems to handle model uncertainty represents the main novel contribution of the present work.

\end{abstract}

\section{Introduction} \label{sec_intro}
Cyber-Physical Systems (CPS) are physical processes that are tightly integrated with computation and communication systems for monitoring and control. Though advances in CPS design has equipped them with adaptability, resiliency, safety, and security features that exceed the simple embedded systems of the past, it often leaves open several points for attackers to strike. CPS security problems have attracted the attention of researchers worldwide recently; some state-of-the-art anomaly detection algorithms can be found in \cite{giraldo_etal, navid_acc_gmm, Pasqualetti_tac}.

A common practice is to model a CPS as either a deterministic system or a stochastic system with additive Gaussian uncertainties. Motivated by the recent developments in distributionally robust optimization (DRO) techniques \cite{dr_goh, dr_peyman, dr_wiesemann}, authors in \cite{venki_lcss, venki_acc, danli_lcss} have developed DRO anomaly detectors that remove assumptions on specific functional forms of the uncertainties in the stochastic CPS model. On the other hand, it is a common practice to assume that the true CPS dynamics are known exactly. Unfortunately, modeling and sampling errors are inherent and significant in working with real systems due to nonlinearities, learned (system identification, machine learning) models, adaptive models, or simply due to changing environmental conditions or aging. A multiplicative noise framework for capturing model uncertainty offers several compelling advantages over additive noise models. It provides a statistical description of the uncertainty that depends on the control input and state \cite{li_acc_multi_estimation, ben_ifac, ben_tac}. Using a multiplicative noise model, however, requires new tools to build and tune anomaly detectors that accommodate the more general functional form of the model.

State estimation is a crucial component in any model-based anomaly detector design, which depends on a state-space model for the system dynamics. This dependency causes limitations on the usage of the classical Kalman filter as it critically relies on the availability of an accurate state-space model, making it susceptible to model risk. Robust Kalman filtering with additive uncertainties was explored in \cite{levy_tac}, where the uncertain joint distribution of the states and outputs was accounted for. Another robust Kalman filter design was developed using a $\tau$-divergence based family of distributions in \cite{zorzi_tac}. In \cite{wdrkf_peyman}, a Wasserstein distributionally robust Kalman filter (W-DR-KF) was developed to account for distributional uncertainty. However, a procedure for jointly computing a pair of state estimator and feedback gain to guarantee stability in this setting remains unexplored. 

Although stochastic modeling of CPS with additive uncertainty is well studied, there are no works to the best of our knowledge which have considered both multiplicative and additive noises together in the CPS security literature. The evolution of non-Gaussian state distributions under the effect of multiplicative noise invalidates use of the standard Kalman filter, as the separation principle available in linear quadratic Gaussian (LQG) setting in \cite{kalman1960contributions} no longer holds. Though \cite{li_acc_multi_estimation} considered both multiplicative and additive noises in an optimal control setting, a restrictive Gaussian assumption was imposed on the uncertainties. The approach in this paper builds on the foundation established by \cite{dekoning_tac}, where the multiplicative noise-driven LQG (MLQG) problem was solved by posing a set of coupled algebraic Riccati equations, from which the optimal linear output feedback controller and estimator gains were jointly computed.

\emph{Contributions:} This paper is part of our ongoing work \cite{venki_lcss, venki_acc} to leverage powerful results in control theory and distributionally robust optimization to design robust anomaly detectors. Specifically, the detector threshold corresponding to a desired false alarm rate in the setting considered in this paper was computed through the moment-based approaches explained \cite{venki_lcss}. In prior work we addressed detectors robust to non-Gaussian additive noise. In this work, 
\begin{enumerate}
    \item We design an anomaly detector for stochastic linear cyber-physical systems that is robust to modeling errors. To our knowledge, this is the first paper to consider tuning an anomaly detector for a system model that incorporates model uncertainty. We propose a multiplicative noise framework and integrate the MLQG compensator to compute the residual.
    \item We demonstrate our proposed approach using numerical simulations and show that multiplicative noises result in greater anomaly detector thresholds as long as mean square compensatability conditions are satisfied.  
\end{enumerate}
The rest of the paper is organized as follows. In \S \ref{sec_wasserstein_problem}, the problem of monitoring an uncertain CPS with model uncertainty is formulated. Then, the multiplicative noise driven LQG compensator is discussed in \S \ref{sec_multi_lqg}. Subsequently, the anomaly detector design is presented in \S \ref{sec_anomaly_detector}. The proposed idea is then demonstrated using a numerical simulation in \S \ref{sec_num_results}. Finally, the paper is closed in \S \ref{sec_conclusion} along with directions for future research. 

\section*{Notations \& Preliminaries}
The set of real numbers, integers are denoted by $\bbr, \bbz$. The subset of real numbers greater than $a \in \bbr$ is denoted by $\bbr_{> a}$.
The set of integers between two values $a,b \in \bbz$ with $a<b$ is denoted by $[a:b]$. 
We denote by $\mathbb{S}^{n}$ the set of symmetric matrices in $\bbr^{n \times n}$ and the cone of positive definite (semi-definite) matrices on $\mathbb{S}^{n}$ as $\mathbb{S}^{n}_{++} (\mathbb{S}^{n}_{+})$. An identity matrix in dimension $n$ is denoted by $I_{n}$. The Kronecker product of two matrices $A \in \bbr^{m \times n}, B \in \bbr^{p \times q}$ is denoted by $A \otimes B$ and the vectorization of a matrix $A \in \bbr^{m \times n}$ is denoted by $\vect(A) \in \bbr^{mn}$ and the matricization of vector $x \in \bbr^{p}$ is denoted by $\text{mat}(x, n, m) \in \bbr^{n \times m}$ where $n \times m = p$. The trace of a matrix $A \in \mathbb{R}^{n \times n}$ is denoted by $\mathbf{Tr}(A)$. A probability distribution with mean $\mu$ and covariance $\Sigma$ is denoted by $\mathbb{P}(\mu, \Sigma)$, and specifically $\mathcal{N}_{d}(\mu, \Sigma)$ if the distribution is normal in $\mathbb{R}^{d}$. Given a matrix $A \in \mathbb{R}^{n \times n}$ and a vector valued random variable $z \in \mathbb{R}^{p}, p \geq 1$ with $\mathbb{E}[z] = \mu, \mathbb{E}[(z-\mu)(z-\mu)^{\top}] = \Sigma$, then $\mathbb{E}[z^{\top} A z] = \mathbf{Tr}(A \Sigma) + \mu^{\top} A \mu$.

\section{Problem Formulation} \label{sec_wasserstein_problem}
\subsection{Uncertain CPS Model}
We model an uncertain CPS for time $k \in \bbn$ using a stochastic discrete-time linear time varying (LTV) system:
\begin{align} 
    x_{k+1} &= {A_{k}} x_{k} + {B_{k}} u_{k} + w_{k}, \label{eqn_uncertain_cps} \\
    y_{k} &= {C_{k}} x_{k} + v_{k}. \label{eqn_cps_output_model}
\end{align}
Here, $x_{k} \in \bbr^n$, $u_{k} \in \bbr^m$, and $y_k \in \bbr^p$ are the system state, control input, and output at time $k$.
The next-state $x_{k+1} \in \bbr^{n}$ is a random linear combination of the current state and process noise $w_{k}$, which is a zero-mean white noise process.
Similarly, the output $y_{k} \in \bbr^{p}$ is a random linear combination of the states and the sensor noise $v_{k}$, which is a zero-mean white noise process.
The initial state is a random variable $x_0 \sim \mathbb{P}_{x_0}(0, \Sigma_{x_{0}})$.
The system matrices are decomposed as
\begin{alignat}{3}
    &  A_k = \left(\Bar{A} + \hat{A}_{k} \right), \
    && B_k = \left(\Bar{B} + \hat{B}_{k} \right), \
    && C_k = \left(\Bar{C} + \hat{C}_{k}\right), \nonumber \\
    &  \hat{A}_{k} = \sum^{n_{a}}_{i=1} \gamma_{ki} \mathcal{A}_{i}, \
    && \hat{B}_{k} = \sum^{n_{b}}_{j=1} \delta_{kj} \mathcal{B}_{j}, \
    && \hat{C}_{k} = \sum^{n_{c}}_{l=1} \kappa_{kl} \mathcal{C}_{l}. \label{eqn_multi_matrices}
\end{alignat}
where $\bar{A}$, $\bar{B}$, $\bar{C}$ denote the nominal dynamics, control, and output matrices respectively. Given the constants, $n_{a}, n_{b}, n_{c} \in \bbz_{>0}$, the multiplicative noise terms are modeled by the i.i.d. across time (white), zero-mean, mutually independent scalar random variables $\gamma_{ki}$, $\delta_{kj}$, $\kappa_{kl}$, which have variances $\sigma^{2}_{a,i}$, $\sigma^{2}_{b,j}$, $\sigma^{2}_{c,l}$ for $i \in [1:n_{a}]$, $j \in [1:n_{b}]$, $l \in [1:n_{c}]$ respectively. 
The pattern matrices $\mathcal{A}_{i} \in \bbr^{n \times n}$, $\mathcal{B}_{j} \in \bbr^{n \times m}$, and $\mathcal{C}_{l} \in \bbr^{p \times n}$ specify how each scalar noise term affects the system matrices. 
It is then evident from \eqref{eqn_uncertain_cps} and \eqref{eqn_cps_output_model} that $\hat{A}_{k}, \hat{B}_{k}$, and $\hat{C}_{k}$ quantify uncertainty about the nominal system matrices $\bar{A}, \bar{B}$, and $\bar{C}$ respectively. The distributions of all the scalar multiplicative noise random variables are assumed to be known. The covariance of the additive noises\footnote{Even when the primitive random variables $w_k, v_k, x_0$ are assumed to be Gaussian, the resulting $\bbp_{x_k}$ at any time step $k > 0$ will be \emph{non-Gaussian} due to the multiplicative noise.} $(\Sigma_{w}, \Sigma_{v})$ are assumed to be known; they may be estimated from collected data via e.g. bootstrap sample averaging. For simplicity, we assume that $x_{0}$ and all the additive, multiplicative noises $w_{k}, v_{k}, \{\gamma_{ki}\}^{n_{a}}_{i=1}, \{\delta_{kj}\}^{n_{b}}_{j=1}, \{\kappa_{kl}\}^{n_{c}}_{l=1}$ are mutually independent of each other. We denote the first moment, second moment, and covariance of the state at time $k$ as $\mu_{x_k} = \bbe\left[ x_{k} \right]$, $V_k = \bbe\left[ x_{k} x_{k}^\top \right] $, and  $\Sigma_{x_k} = \bbe\left[ (x_{k} - \mu_{x_k}) (x_{k}-\mu_{x_k})^\top \right]$, respectively. Likewise, we denote the first moment, second moment, and covariance of the output at time $k$ as $\mu_{y_k} = \bbe\left[ y_{k} \right]$, $Y_k = \bbe\left[ y_{k} y_{k}^\top \right] $, and  $\Sigma_{y_k} = \bbe\left[ (y_{k} - \mu_{y_k}) (y_{k}-\mu_{y_k})^\top \right]$, respectively. 

\subsection{Review of Concepts}
Here, we re-state some definitions from \cite{dekoning_tac} on the mean squared versions of stabilizability, detectability and the resulting compensatability of systems given by \eqref{eqn_uncertain_cps} and \eqref{eqn_cps_output_model}. 

\textbf{Definition 1:}
The system in \eqref{eqn_uncertain_cps} is \emph{mean-square stable} if $\forall x_{0} \in \bbr^{n}, \exists V_{\infty} \in \bbs^{n}_{+}$ such that
\begin{align*}
\lim_{k \rightarrow \infty} V_{k} = \lim_{k \rightarrow \infty} \bbe\left[ x_{k} x^{\top}_{k} \right] \rightarrow V_{\infty}.
\end{align*}

\textbf{Definition 2:} 
The system in \eqref{eqn_uncertain_cps} is \emph{mean-square stabilizable} if there exists a control gain matrix $K \in \bbr^{m \times n}$ such that using controls $u_k = K x_k$ makes \eqref{eqn_uncertain_cps} mean-square stable.

\textbf{Definition 3:}
The system in \eqref{eqn_uncertain_cps} and \eqref{eqn_cps_output_model} is \emph{mean-square compensatable} if there exist control and filter gain matrices $K \in \bbr^{m \times n}$ and $L \in \bbr^{n \times p}$ such that the system 
\begin{align*}
    \begin{bmatrix}
    x_{k+1} \\ \hat{x}_{k+1}
    \end{bmatrix}
    =
    \begin{bmatrix} A_{k} & B_{k}K \\ LC_{k} & \bar{A}+\bar{B}K - L \bar{C} \end{bmatrix}
    \begin{bmatrix}
    x_{k} \\ \hat{x}_{k}
    \end{bmatrix}
\end{align*}
is mean-square stable.

\textbf{Assumptions}
\begin{enumerate}
    \item The system given by \eqref{eqn_uncertain_cps} and \eqref{eqn_cps_output_model} is \emph{mean-square compensatable}.
    \item The optimal state estimator at any time $k$ given \eqref{eqn_uncertain_cps} and \eqref{eqn_cps_output_model} is an affine\footnote{It is possible to design a nonlinear state estimator to outperform a given affine estimator in this setting. However, it is out of the scope of this paper.} function of the output $y_{k}$.
\end{enumerate}

\begin{problem}
Under the above assumptions for a given stochastic CPS model specified by \eqref{eqn_uncertain_cps}, \eqref{eqn_cps_output_model}, obtain residual data from an appropriate state estimator module that accounts for both multiplicative and additive noises, and subsequently design an anomaly detector threshold such that the worst case false alarm rate does not exceed a desired value. 
\end{problem}


\section{Residuals via Multiplicative Noise LQG} \label{sec_multi_lqg}
Due to the multiplicative noises in \eqref{eqn_uncertain_cps} and \eqref{eqn_cps_output_model}, the state distribution will be non-Gaussian even when all primitive noise distributions are Gaussian. Further, the classical separation principle from the additive noise setting does not hold in presence of multiplicative noises \cite{dekoning_tac}. This necessitates a framework where the optimal controller and the estimator gains are computed \emph{jointly}. Here, we elaborate on obtaining the residual from CPS using the multiplicative noise-driven LQG and show that the residual covariance is a function of both additive and multiplicative noise covariance matrices.

\subsection{Designing Multiplicative Noise-Driven LQG}
Under both multiplicative and additive noises in the system, the optimal linear output feedback controller can be exactly computed through the combination of a multiplicative noise KF with a multiplicative noise LQR as described in \cite{el_ghaoui_lcss, dekoning_tac, li_acc_multi_estimation}. We consider the multiplicative noise-driven linear-quadratic Gaussian (MLQG) optimal control problem, which requires finding an output feedback controller $u_{k} = \pi_{k}(y_{0:k})$ for a system given by \eqref{eqn_uncertain_cps} and \eqref{eqn_cps_output_model}:
\begin{equation}
    \begin{aligned}
    &\underset{\pi_{k} \in \Pi_{k}}{\text{minimize}} \quad \lim_{T \rightarrow \infty} \frac{1}{T} \bbe_{\mathcal{E}_{k}}\left[\sum^{T-1}_{k=0} x^{\top}_{k} Q x_{k} + u^{\top}_{k} R u_{k} \right], \\
    &\text{subject to} \quad  \eqref{eqn_uncertain_cps}, \eqref{eqn_cps_output_model},
    \end{aligned}
\end{equation}
where $\mathcal{E}_{k} = \left\{x_{0}, \{\hat{A}_{k}\}, \{\hat{B}_{k}\},\{\hat{C}_{k}\},\{w_{k}\},\{v_{k}\}\right\}$, $Q \succeq 0$, $R \succ 0$. Then, the optimal linear compensator gain matrices can be computed by solving the following coupled nonlinear matrix Riccati equations in symmetric matrix variables $P_1,P_2,P_3,P_4 \in \bbs^{n}_{+}$:
\begin{align}
    P_{1} &= Q + \bar{A}^{\top} P_{1} \bar{A} + \sum^{n_{a}}_{i = 1} \sigma^{2}_{a,i} \mathcal{A}^{\top}_{i} P_{1} \mathcal{A}_{i} - K^{\top} K_{\alpha} K \nonumber \\
    &+ \sum^{n_{a}}_{i = 1} \sigma^{2}_{a,i} \mathcal{A}^{\top}_{i} P_{2} \mathcal{A}_{i} + \sum^{n_{c}}_{i = 1} \sigma^{2}_{c,i} \mathcal{C}^{\top}_{i} L^{\top} P_{2} L \mathcal{C}_{i}, \\
    P_{2} &= (\bar{A} - L \bar{C})^{\top} P_{2} (\bar{A} - L \bar{C}) + K^{\top} K_{\alpha} K, \\
    P_{3} &= \Sigma_{w} + \bar{A} P_{3} \bar{A}^{\top} - L L_{\alpha} L^{\top} + \sum^{n_{a}}_{i = 1} \sigma^{2}_{a,i} \mathcal{A}_{i} P_{3} \mathcal{A}^{\top}_{i}  \nonumber \\ &+ \sum^{n_{a}}_{i = 1} \sigma^{2}_{a,i} \mathcal{A}_{i} P_{4} \mathcal{A}^{\top}_{i} + \sum^{n_{b}}_{i = 1} \sigma_{b,i}^2 \mathcal{B}_{i} K P_{4} K^{\top} \mathcal{B}^{\top}_{i}, \\
    P_{4} &= (\bar{A}+\bar{B}K) P_{4} (\bar{A}+\bar{B}K)^{\top} + L L_{\alpha} L^{\top},
\end{align}
where for notation simplicity, we denote
\begin{align}
    K_{\alpha} &= R + \bar{B}^{\top} P_{1} \bar{B} + \sum^{n_{b}}_{j=1} \sigma^{2}_{b,j} \mathcal{B}^{\top}_{j} P_{1} \mathcal{B}_{j} + \sum^{n_{b}}_{j=1} \sigma^{2}_{b,j} \mathcal{B}^{\top}_{j} P_{2} \mathcal{B}_{j} \\
    L_{\alpha} &= \Sigma_{v} + \bar{C} P_{3} \bar{C}^{\top} + \sum^{n_{c}}_{j=1} \sigma^{2}_{c,j} \mathcal{C}_{j} P_{3} \mathcal{C}^{\top}_{j} + \sum^{n_{c}}_{j=1} \sigma^{2}_{c,j} \mathcal{C}_{j} P_{4} \mathcal{C}^{\top}_{j}.
\end{align} 
Then, the associated optimal controller and estimator gains $(K, L)$ are given by
\begin{align}   
    K &= -K_{\alpha}^{-1} \bar{B}^{\top} P_{1} \bar{A}, \label{eqn_multi_noise_K}\\
    L &= \bar{A} P_{3} \bar{C}^{\top} L_{\alpha}^{-1}. \label{eqn_multi_noise_L}
\end{align}
Finally, the optimal linear compensator is
\begin{align} 
    u_k &= K \hat{x}_{k}, \quad \text{and}\\
    \hat{x}_{k+1} &= (\Bar{A} + \Bar{B} K) \hat{x}_{k} + L(y_{k} - \Bar{C} \hat{x}_{k}), \nonumber \\
    &= (\Bar{A} + \Bar{B} K - L \bar{C}) \hat{x}_{k} + L C_{k} x_{k} + L v_{k} \label{eqn_xhat_multi_lqg}
\end{align}
It is necessary to account for the multiplicative noise to achieve the minimum quadratic cost; furthermore, it is straightforward to find systems in \eqref{eqn_uncertain_cps} and \eqref{eqn_cps_output_model} which are \emph{mean-square unstable} when controlled by (multiplicative-noise-ignorant) LQG, meaning that it is necessary to account for multiplicative noise to achieve mean-square stability.

\subsection{Residual from Multiplicative Noise LQG}
We define the estimation error as $e_{k} = x_{k} - \hat{x}_{k}$. Then the estimation error evolves as follows
\begin{equation}
    \footnotesize e_{k+1} = (\bar{A} - \hat{B}_{k} K - L \bar{C}) e_{k} + (\hat{A}_{k} + \hat{B}_{k} K - L \hat{C}_{k}) x_{k} + w_{k} - L v_{k}.    
\end{equation}
It is evident from above that estimation error is a function of the multiplicative noise terms. We now elaborate how to obtain the residual signal required for anomaly detection. Define the residual $r_{k} \in \bbr^{p}$ as
\begin{align} 
    r_{k} &= y_{k} - \bar{C} \hat{x}_{k} = \bar{C} e_{k} + \hat{C}_{k} x_{k} + v_{k} \quad \text{and} \label{eqn_r_multi_lqg} \\
    \bbe [r_{k}]
    &= \bbe [\bar{C} e_{k} + \hat{C}_{k} x_{k} + v_{k}] 
    = \Bar{C} \bbe [e_k]. \label{eq:first_moment_output_residual}
\end{align}
\begin{figure*}
\begin{equation*}
\small 
    H = 
    \begin{bmatrix}
    \bar{A} \otimes \bar{A} + \Sigma_A^\prime \! & \!(\bar{B} K) \otimes \bar{A}\! & \!\bar{A} \otimes (\bar{B} K)\! & \!\left( \bar{B} \otimes \bar{B} + \Sigma_B^\prime \right) (K \otimes K) \\
    (L \Bar{C}) \otimes \Bar{A}\! & \!(\Bar{A} + \Bar{B} K - L \Bar{C}) \otimes \Bar{A}\! & \!(L \Bar{C}) \otimes (\Bar{B} K)\! & \!(\Bar{A} + \Bar{B} K - L \Bar{C}) \otimes (\Bar{B} K) \\
    \Bar{A} \otimes (L \Bar{C})\! & \!(\Bar{B} K) \otimes (L \Bar{C})\! & \!\Bar{A} \otimes (\Bar{A} + \Bar{B} K - L \Bar{C})\! & \!(\Bar{B} K) \otimes (\Bar{A} + \Bar{B} K - L \Bar{C}) \\
    (L \otimes L) (\bar{C} \otimes \bar{C} + \Sigma_C^\prime)\! & \!(\Bar{A} + \Bar{B} K - L \Bar{C} ) \otimes (L \Bar{C})\! & \!(L \Bar{C}) \otimes (\Bar{A} + \Bar{B} K - L \Bar{C} )\! & \!(\Bar{A} + \Bar{B} K - L \Bar{C} ) \otimes (\Bar{A} + \Bar{B} K - L \Bar{C} )
    \end{bmatrix}.
\end{equation*}
\caption{The Matrix $H$ in \eqref{eqn_state_second_moment_compact} with terms containing the second moments of entries of the vector $\mathcal{X}_k$.}
\end{figure*}
\noindent Then, $r_{k}$ is not necessarily Gaussian due to the multiplicative noise and has mean $\mathbb{E}[r_{k}] = \bar{C} \mathbb{E}[e_{k}]$ (it becomes zero mean $\forall k \geq 0$ if $e_{0} = 0$) with raw second moment matrix whose vectorized form is given by
\begin{align} \label{eqn_cov_r_multi_lqg}
    R_{k} &= \left( \bar{C} \otimes \bar{C} \right) E_{k} + \bbe\left[ \hat{C}_{k} \otimes \hat{C}_{k} \right] X_{k} +  \vect\left(\Sigma_{v}\right).
\end{align}
To compute the steady state raw second moments of the residual $r_{k}$, we define
\begin{align*}
    E_{k} &= \vect\left(\bbe[e_{k}e^{\top}_{k}]\right), \quad X_{k} = \vect\left(\bbe[x_{k}x^{\top}_{k}]\right), \\
    \tilde{X}_k &= \vect \left(\bbe \left[ x_{k} \hat{x}_{k}^\top \right]\right), \quad \breve{X}_k = \vect \left(\bbe \left[ \hat{x}_{k} x_{k}^\top \right]\right), \\ \hat{X}_k &= \vect \left(\bbe \left[ \hat{x}_{k} \hat{x}_{k}^\top \right]\right), \quad R_{k} = \vect\left(\bbe[r_{k}r^{\top}_{k}]\right)
\end{align*}
\begin{align*}
    \mathcal{X}_k &\Let 
    \begin{bmatrix}
    X^{\top}_k &
    \tilde{X}^{\top}_k &
    \breve{X}^{\top}_k &
    \hat{X}^{\top}_k
    \end{bmatrix}^{\top}, \quad
    \mathcal{V}
    \Let
    \begin{bmatrix}
    \vect (\Sigma_w) \\
    \vect (\Sigma_v)
    \end{bmatrix}, \\
    \Sigma_A^\prime &= \bbe \left[ \hat{A}_k \otimes \hat{A}_k \right] = \sum_{i=1}^{n_a} \sigma^2_{a,i} (\mathcal{A}_i \otimes \mathcal{A}_i), \\
    \Sigma_B^\prime &= \bbe \left[ \hat{B}_k \otimes \hat{B}_k \right] = \sum_{j=1}^{n_b} \sigma^2_{b,j} (\mathcal{B}_j \otimes \mathcal{B}_j), \\
    \Sigma_C^\prime &= \bbe \left[ \hat{C}_k \otimes \hat{C}_k \right] = \sum_{l=1}^{n_c} \sigma^2_{c,l} (\mathcal{C}_l \otimes \mathcal{C}_l).
\end{align*}
Then, it is straight forward to see that $\mathcal{X}_k$ evolves as follows
\begin{align} \label{eqn_state_second_moment_compact}
    \mathcal{X}_{k+1} = H \mathcal{X}_k + \underbrace{\begin{bmatrix}
    I_{n} \otimes I_{n} & 0_{n^{2} \times 1} \\
    0_{n^{2} \times n^{2}} & 0_{n^{2} \times 1} \\
    0_{n^{2} \times n^{2}} & 0_{n^{2} \times 1} \\
    0_{n^{2} \times n^{2}} & L \otimes L
    \end{bmatrix}}_{:=\Phi} \mathcal{V},
\end{align}
where the matrix $H$ in \eqref{eqn_state_second_moment_compact} gathers all the resulting coefficients obtained while expanding the entries of the vector $\mathcal{X}_k$. The algebra resulting in the following expression of $H$ is available in the appendix of \cite{venki_multi_noise_anomaly_arxiv}. Since the optimal gain matrices $K,L$ achieve mean-square compensation of the system \eqref{eqn_uncertain_cps} and \eqref{eqn_cps_output_model}, the covariance of the estimation error will have a steady state value. Since by assumption, $\Bar{A} - L \Bar{C}$ is Schur stable, we see that $\bbe [e_k] \to 0$ as $k \to \infty$ regardless of the initial state-residual $e_0$ which in turn results in $\bbe [r_k] \to 0$ as $k \to \infty$. That is, $    \bbe[e_\infty] = 0 \implies \bbe[r_\infty] = 0$
and subsequently in steady state,
\begin{align}
    \mathcal{X}_{\infty} &= H \mathcal{X}_\infty + \Phi \mathcal{V}. \\
    \iff \mathcal{X}_{\infty} &= (I_{4n^2} - H)^{-1} \Phi \mathcal{V}.
\end{align}
This amounts to solving a (generalized) Lyapunov equation. Such an equation can be solved more efficiently by specialized solvers which do not require the inverse to be computed explicitly; for simplicity we present the equation and its solution in this form. However, the Schur stability of the matrix $H$ subject to the mean-square compensation achieved by the matrices $(K,L)$ determines whether the resulting $\mathcal{X}_{\infty}$ (which exists no matter whatever approach is used to compute it) can be employed to compute the steady state residual moments. For instance, in a strong multiplicative noise setting, the matrix $H$ defined using $(K,L)$ matrices that do not achieve mean-square compensation will \emph{not} be Schur stable and the resulting $\mathcal{X}_{\infty}$ cannot be used meaning that steady state $\Sigma_r$ does not exist. Having obtained a valid $\mathcal{X}_{\infty}$, the steady state second moments of the state- and output-residuals can then be computed as
\begin{align}
    E_{\infty} &= X_\infty - \tilde{X}_\infty - \breve{X}_\infty + \hat{X}_\infty, \quad \text{and} \\
    R_{\infty} &= (\Bar{C} \otimes \Bar{C}) E_\infty + \Sigma_C^\prime X_\infty + \vect(\Sigma_v).
\end{align}
Finally, using the matrix reshaping operator $\text{mat}(\cdot)$, we retrieve the steady state $\Sigma_r$ as follows
\begin{align}
\Sigma_{x_\infty} &= \text{mat}(E_{\infty}, n, n), \quad \text{and} \\ \Sigma_r &= \text{mat}(R_{\infty}, p, p).    
\end{align}

\section{Anomaly Detector Design with Residual from MLQG Compensation}\label{sec_anomaly_detector}
We now present how to analyze the residual obtained from the MLQG compensator and elaborate the procedure to construct the corresponding anomaly detector threshold in this section. Note that the covariance of the residual computed through \eqref{eqn_cov_r_multi_lqg} is a function of covariance matrices of both the additive and multiplicative noises. This is in sharp contrast to the case in \cite{venki_lcss}, \cite{venki_acc} where the residual covariance was just a function of the additive noise covariance. Further, to account for the changes in the covariance of the residual, we form a quadratic distance measure as  
\begin{align} \label{eqn_q}
q_{k} = r^{\top}_{k} \Sigma^{-1}_{r} r_{k}.
\end{align}
It is then straightforward to see that
\begin{align} \label{eqn_Eofq}
\mathbb{E}[q_{k}] 
&= \mathbb{E}[r^{\top}_{k} \Sigma^{-1}_{r} r_{k}] \nonumber \\
&= \mathbf{Tr}(\Sigma^{-1}_{r} \Sigma_{r}) + (\bar{C} \mathbb{E}[e_{k}])^{\top} \Sigma^{-1}_{r} (\bar{C} \mathbb{E}[e_{k}])  \nonumber  \\
&= p + (\mathbb{E}[e_{k}])^{\top} \bar{C}^{\top} \Sigma^{-1}_{r} \bar{C} \mathbb{E}[e_{k}].
\end{align}
This implies that \eqref{eqn_Eofq} is applicable only when mean-square compensation is achieved through properly designed $(K, L)$ matrix pair as the steady state $\Sigma_r$ is guaranteed to exist in that case. Then, for a given $q_{k}$ from \eqref{eqn_q} and a threshold $\alpha \in \bbr_{>0}$ corresponding to a desired false alarm rate $\mathcal{F}$, the anomaly detector can be designed such that alarm time(s) $k^{\star} \in \mathbb{N}$ are produced according to the following rules
\begin{align} \label{eqn_detector_thresh}
    \begin{cases} q_{k} \leq \alpha, &\text{no alarm}, \\
    q_{k} > \alpha, &\text{alarm: } k^{\star} = k.
    \end{cases}
\end{align}
If $\mathbb{P}_{r_{k}}$ was Gaussian, then $q_{k}$ would follow the chi-squared distribution, meaning that for a given tail probability defined using $\mathcal{F}$, the chi-squared detector described as in \cite{navid_acc_gmm} can be used to obtain the required detector threshold. However, in our setting due to the multiplicative noises, $\mathbb{P}_{r_{k}}$ is \emph{non-Gaussian} and thereby the chi-squared detector is \emph{not} appropriate. We instead utilize a moment-based approach for constructing the threshold. We propose to use the higher-order moment based anomaly detector design proposed in \cite{venki_lcss} to design the detector threshold in this setting. The residual $q_{k}$ is collected for a sufficiently long period of time to form the $s$-moments based ambiguity set $\mathcal{P}^{s}_{q} := \left\{ \bbp_{q} \mid \bbe[q^{s}_{k}] = M^{s}_{q} \right\}$. The optimal threshold $\alpha^{\star}_{q,s}$\footnote{The two subscripts $q,s$ in $\alpha^{\star}_{q,s}$ denote the random variable and the number of moments considered respectively.} satisfying
\begin{align} \label{eqn_s_moment_bd_prob}
    \sup_{\bbp_{q} \in \mathcal{P}^{s}_{q}} \bbp_{q} \left[q_{k} > \alpha^{\star}_{q,s} \right] \leq \mathcal{F},
\end{align}
can then be obtained by directly invoking Theorem 4 in \cite{venki_lcss} corresponding to a given desired false alarm rate $\mathcal{F}$. 

\section{Numerical Results} \label{sec_num_results}
We consider an inverted pendulum with a torque-producing actuator whose dynamics have been linearized about the vertical equilibrium. That is, the pendulum of mass $m$ is suspended by a mass-less rod of length $l$ and the angle $\theta$ is measured from the downward vertical with positive counter clockwise direction. The corresponding nonlinear differential equation of the pendulum mass is 
\begin{align} \label{eqn_pendulum_nonlinear}
    \Ddot{\theta} = m_{c} \sin(\theta) + \tau, 
\end{align}
where $m_{c} = -\frac{g}{l}$ denotes the uncertain mass constant. Let us denote the state vector by $x = \begin{bmatrix}x_{1} & x_{2}\end{bmatrix} = \begin{bmatrix}\theta & \dot{\theta}\end{bmatrix}$ and the torque input by $u = \tau$. Then, the corresponding discrete time dynamics obtained through the forward Euler discretization of the linearized dynamics of \eqref{eqn_pendulum_nonlinear} around the equilibrium point $\Tilde{x} = (\pi, 0)$ with step size $\Delta t$ is 
\begin{align} \label{eqn_x_sim}
    x_{k+1} = \begin{bmatrix} 1 & \Delta t \\ m_{c} \Delta t & 1 \end{bmatrix} x_{k} + \begin{bmatrix}0 \\ \Delta t \end{bmatrix} u_{k} + w_{k}. 
\end{align}
Uncertainty on the mass constant $m_{c}$ corresponds to uncertainty on the matrix A. We consider an example where the true mass constant is $m_{c}$ = 10, but the nominal model underestimates it as $m_{c} = 5$. We take a step size $\Delta t = 0.1$. At discrete time instances, the sensor returns a noisy measurement of the angular position of pendulum. Hence the corresponding linearized noisy output model is, 
\begin{align} \label{eqn_y_sim}
    y = \theta + v_{k} = \begin{bmatrix} 1 & 0 \end{bmatrix} x_{k} + v_{k}. 
\end{align}
Both $w_{k}$ and $v_{k}$ are sampled from the multivariate Laplacian (which has heavier tails than Gaussian with same mean and covariance) with zero-mean and covariance $\Sigma_{w} = 2I_{n}$, $\Sigma_{v} = 2I_{p}$ respectively. The state and control penalty matrices are $Q = I_{n}, R  = I_{m}$ respectively. The multiplicative noise was considered to exist both in the $A$ and $C$ matrices, with the direction matrices being $\mathcal{A}_{1} = \begin{bmatrix} 0 & 0 \\ 1 & 0\end{bmatrix}$ and $\mathcal{C}_{1} = \begin{bmatrix} 0.1 & 0 \end{bmatrix}$ with the multiplicative noise variances $\gamma_{k,1} \sim \mathcal{N}(0, \sigma^{2}_{a,1}), \kappa_{k,1} \sim \mathcal{N}(0, \sigma^{2}_{c,1})$ respectively. The non-Gaussian additive primitive noises $w_{k}, v_{k}$ along with these multiplicative noises render the traditional chi-squared detector to be ineffective as the system states will evolve to be \emph{non-Gaussian} for all $t  > 0$. Through simulation, we collected the quadratic distance measure $q_k$ data for $T = 10^{7}$ time steps for the above system with multiplicative noises under two different settings namely, 1) using the standard LQG, and 2) using multiplicative noise-driven LQG compensators. The $q_k$ data was then used to tune the anomaly detector for a desired false alarm rate of $\mathcal{F} = 5\%$ using Theorem 4 in \cite{venki_lcss} with $s = 4$ moments in \eqref{eqn_s_moment_bd_prob} and along with a bisection tolerance of $\epsilon = 10^{-4}$. The resulting moment bound problem was solved using the SOSToolbox on MATLAB with the SeDuMi solver. The code is made publicly available at \url{https://github.com/TSummersLab/AnomalyDetectionMultiplicativeNoise}
\begin{figure}
    \centering
    \includegraphics[width=\linewidth]{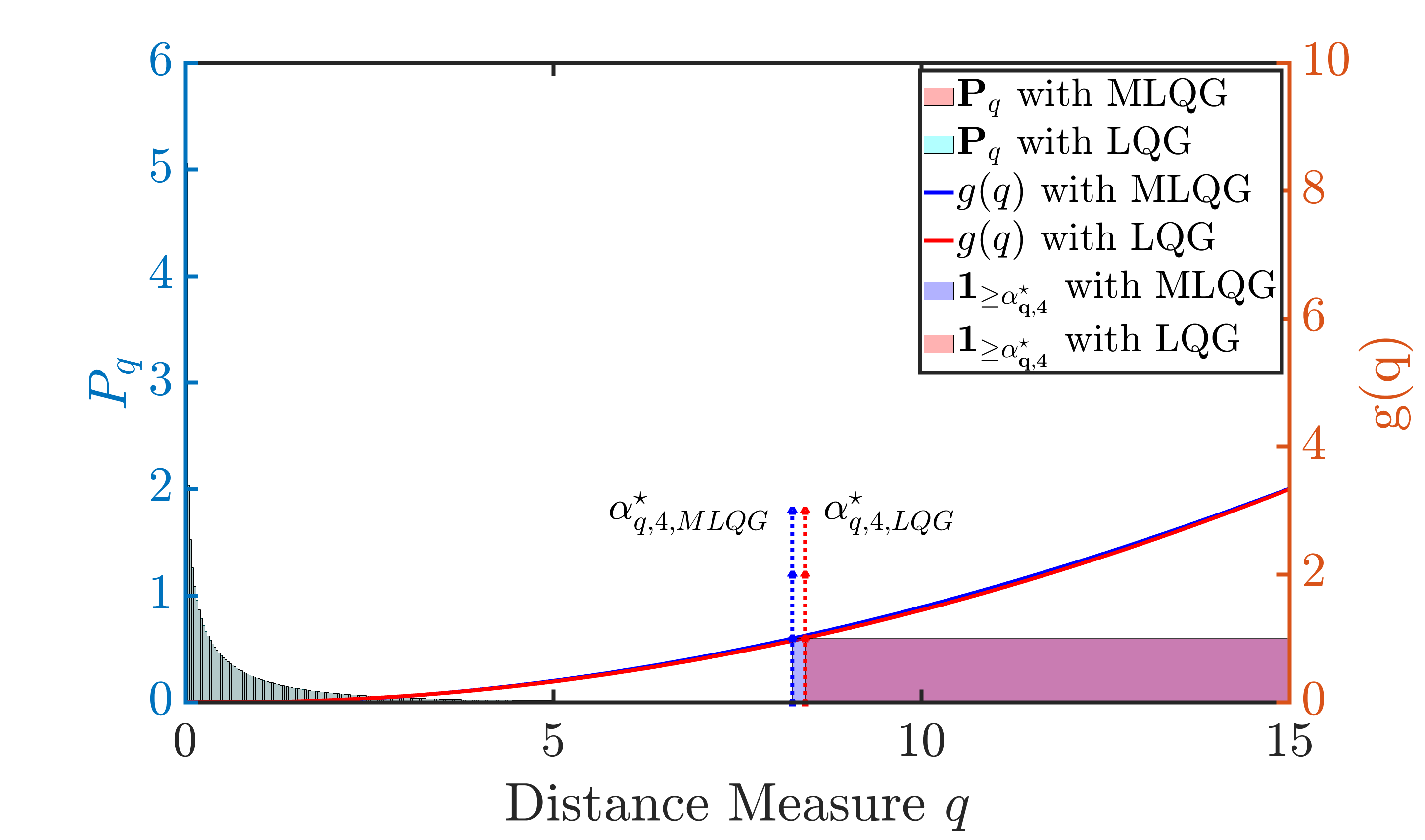}
    \caption{\textbf{Detector Threshold With Multiplicative Noise:} The histograms of the $q_{k}$ using the MLQG and LQG estimators with $\sigma^{2}_{a,1} = \sigma^{2}_{c,1} = 0.06$ are shown in red and cyan colors respectively. The moment based polynomial $g(q)$ shown in blue and red curves bound their indicator functions in shaded blue and red respectively. Though both MLQG and LQG achieve mean-square compensation, the MLQG results in a tighter threshold than the LQG.}
    \label{fig_sim_indicate}
\end{figure}

\subsection{LQG \& MLQG with Low Multiplicative Noises}
\begin{table}[h]
    \centering
{\setlength{\tabcolsep}{5pt}
\begin{tabular}{|c|c|c|c|c|c|}
\hline
\multirow{2}{*}{Method} &
\multicolumn{5}{c|}{$\sigma^{2}_{a,1} = \sigma^{2}_{c,1}$} \\
\cline{2-6}
 & \scriptsize 0.02 & \scriptsize 0.04 & \scriptsize 0.06 & \scriptsize 0.08 & \scriptsize 0.10 \\
\hline \hline
$\lambda_{\max}(H)$ LQG & 0.9105 & 0.9414 & 0.9625 & 0.9789 & 0.9926\\
\hline
$\lambda_{\max}(H)$ MLQG & 0.8908 & 0.9071 & 0.9159 & 0.9217 & 0.9259 \\
\hline
\end{tabular}
}
\caption{Effect of varying the low multiplicative noise variances $(\sigma^{2}_{a,1} = \sigma^{2}_{c,1})$ on the resulting $\lambda_{\max}(H)$ corresponding to LQG and MLQG compensators are shown here.}
\label{tab_low_multi_noise_vars}
\end{table}
When the system was simulated with low multiplicative noise variances $\sigma^{2}_{a,1} = \sigma^{2}_{c,1} \leq 0.10$, the resulting $(K,L)$ matrix pair from both the LQG and the MLQG compensators had similar values and the anomaly detectors from both compensators had similar good performances. However, the performance of MLQG started getting better with $\sigma^{2}_{a,1} = \sigma^{2}_{c,1} > 0.10$ and the results with $\sigma^{2}_{a,1} = \sigma^{2}_{c,1} = 0.06$ are shown in Figure \ref{fig_sim_indicate}. The histograms of the $q_{k}$ data using the MLQG and LQG estimators are shown in red and cyan colors respectively. The mean-square compensation of the MLQG compensator was verified via the convergence of the coupled Riccati equations and subsequently the corresponding collected $q_{k}$ data resulted in an optimal detector threshold $\alpha^{\star}_{q,4} = 8.247$ with false alarm rate being $0.89\%$. Similarly, when the $q_{k}$ data collected from the standard LQG was evaluated against a similarly computed threshold $\alpha^{\star}_{q,4} = 8.422$, it resulted in $0.86\%$ false alarms. Though both MLQG and LQG achieve mean-square compensation at a lower noise setting, the MLQG results in a tighter threshold than the LQG. Further, the resulting $H$ matrix from LQG compensator \emph{ceased} to be Schur stable for $\sigma^{2}_{a,1} = \sigma^{2}_{c,1} > 0.11$ agreeing with results in Table \ref{tab_low_multi_noise_vars}. Supposedly, if we used the unstable $H$ matrix in the LQG case, it resulted in $\hat{\mathbb{E}}[q_{k}] \rightarrow \infty$ when the variances became stronger and thereby restricted us from using even the simplest Markov bound in this case to obtain the detector threshold. 

\subsection{Effect of Multiplicative Noise Variance on the Worst Case False Alarm Rate}
\begin{table}[h]
    \centering
{\setlength{\tabcolsep}{5pt}
\begin{tabular}{|c|c|c|c|c|c|c|c|c|}
\hline
$\sigma^{2}_{a,1} = \sigma^{2}_{c,1}$ &
$\lambda_{\max}(H)$ &
$\Sigma_{r}$ &
$\hat{\mathbb{E}}[q_{k}]$ &  
$\mathcal{F}_{worse} (\%)$ & 
$\alpha^{\star}_{q,4}$ \\
\hline \hline
0.15 & 0.9329 & 6.54 & 1.000 & 0.88 & 8.31\\
\hline
0.20 & 0.9372 & 6.73 & 1.000 & 0.87 & 8.37 \\
\hline
0.25 & 0.9403 & 6.92 & 1.000 & 0.79 & 8.67 \\
\hline
0.30 & 0.9426 & 7.10 & 1.000 & 0.74 & 8.91 \\
\hline
\vdots & \vdots & \vdots & \vdots & \vdots & \vdots \\
\hline
3.77 & N/A & N/A & N/A & N/A & N/A \\
\hline
\end{tabular}
}
\caption{Effect of varying the multiplicative noise variances $(\sigma^{2}_{a,1} = \sigma^{2}_{c,1})$ on the resulting $\lambda_{\max}(H)$, steady state $\Sigma_{r}$, optimal threshold $\alpha^{\star}_{q,4}$, sample based mean $\hat{\mathbb{E}}[q_{k}]$ and the worst case false alarm rates $\mathcal{F}_{worse}$ from the MLQG compensator are shown here. It is evident that the MLQG is capable of mean-square compensating the system even with increasing multiplicative noise variances up to a limit.}
\label{tab_multi_noise_vars}
\end{table}
Here, we show how the variances $\sigma^{2}_{a,1}, \sigma^{2}_{c,1}$ of the multiplicative noises $\gamma_{k,1}, \kappa_{k,1}$ respectively affect the resulting anomaly detector's worst case false alarm rate. Starting from $\sigma^{2}_{a,1} = \sigma^{2}_{c,1} = 0.15$, we simulated the system by increasing the variances and the results are in Table \ref{tab_multi_noise_vars}. It is evident that MLQG compensator was capable of mean-square compensate the system with increasing covariances by resulting in finite mean (equal to 1 and thereby agreeing with \eqref{eqn_Eofq}). Starting from $\sigma^{2}_{a,1} = \sigma^{2}_{c,1} \geq 0.45$, numerical issues started accompanying the threshold calculations due to exploding values of the moments (can be addressed using orthogonal basis such as the Legendre polynomial basis to provide numerical stability). Specifically, when the variances were increased beyond $\sigma^{2}_{a,1} = \sigma^{2}_{c,1} \geq 3.77$, the coupled Riccatti equations corresponding to the MLQG stopped converging as mean-square compensation was lost for such higher variance multiplicative noises.
The effect of increasing variance also affected the resulting false alarm rates when the residuals from the MLQG compensator was compared against its respective threshold.
The resulting optimal threshold $\alpha^{\star}_{q,4}$ increased when the multiplicative noise variances increased. For this reason, in this problem setting the false alarm rate of MLQG happened to decrease with increased multiplicative noise variance; there is a nontrivial relation between the multiplicative noise variances and the threshold designed by the detection scheme, which depends e.g. on the coupled Riccati equation solution.
As shown in Table \ref{tab_multi_noise_vars}, the MLQG with finite set of $s = 4$ empirical moments starting from $\hat{\mathbb{E}}[q_{k}]$ guaranteed that the resulting worst case false alarm rate are always upper bounded by the desired value of $\mathcal{F} = 5\%$. 

\section{Conclusion} \label{sec_conclusion}
An extension of the state-of-the-art anomaly detection algorithms for CPS with modeling errors via the multiplicative noise framework was discussed in this paper. The multiplicative noise-driven LQG being a robust state estimator was used to hedge against the model risk to construct the state estimate. The proposed method was demonstrated using a numerical simulation. Future work seeks to investigate the setting where the multiplicative noise distributions are unknown and to obtain online estimates of the system dynamics through system identification technique combined with the above compensator for implementing data-driven distributionally robust anomaly detection for vulnerable CPS.  


\bibliographystyle{IEEEtran}
\bibliography{bibliograph}

\clearpage
\onecolumn
\appendices
\section{Moment dynamics}
Recall the closed-loop system equations:
\begin{align*}
    x_{k+1} &= A_k x_k + B_k u_k + w_k, \\
    \hat{x}_{k+1} &= \Bar{A} \hat{x}_{k} + \Bar{B} u_k + L (y_k - \hat{y}_k), \\
    u_k &= K \hat{x}_k, \\
    y_k &= C_k x_k + v_k, \\
    \hat{y}_k &= \bar{C} \hat{x}_k,
\end{align*}
and the state- and output-residuals
\begin{align*}
    e_k = x_k - \hat{x}_k, \\
    r_k = y_k - \hat{y}_k.
\end{align*}
Denote
\begin{align*}
    \Sigma_A^\prime &= \bbe \left[ \hat{A}_k \otimes \hat{A}_k \right] = \sum_{i=1}^{n_a} \sigma^2_{a,i} (\mathcal{A}_i \otimes \mathcal{A}_i), \\
    \Sigma_B^\prime &= \bbe \left[ \hat{B}_k \otimes \hat{B}_k \right] = \sum_{j=1}^{n_b} \sigma^2_{b,j} (\mathcal{B}_j \otimes \mathcal{B}_j), \\
    \Sigma_C^\prime &= \bbe \left[ \hat{C}_k \otimes \hat{C}_k \right] = \sum_{l=1}^{n_c} \sigma^2_{c,l} (\mathcal{C}_l \otimes \mathcal{C}_l).
\end{align*}
Hence, we have the identities
\begin{align*}
    \bbe \left[ A_k \otimes A_k \right] = \bar{A} \otimes \bar{A} + \Sigma_A^\prime, \\
    \bbe \left[ B_k \otimes B_k \right] = \bar{B} \otimes \bar{B} + \Sigma_B^\prime, \\
    \bbe \left[ C_k \otimes C_k \right] = \bar{C} \otimes \bar{C} + \Sigma_C^\prime.
\end{align*}
While studying the moment dynamics, we shall readily employ the zero-mean and zero-correlation assumptions of $\hat{A}_k$, $\hat{B}_k$, $\hat{C}_k$, $w_k$, and $v_k$ in the following derivations.

\subsection{First moment dynamics}
The expected output-residual is
\begin{align}
    \bbe [r_{k}]
    = \bbe [y_k - \hat{y}_k] 
    = \bbe [C_k x_k + v_k - \Bar{C} \hat{x}_k]
    = \bbe [C_k x_k] - \Bar{C} \bbe [\hat{x}_k] = \Bar{C} \bbe [ x_k - \hat{x}_k] = \Bar{C} \mathbb{E}[e_k]. \label{eq:first_moment_output_residual}
\end{align}
The expected state-residual evolves as
\begin{align}
    \bbe [e_{k+1}]
    &= \bbe[ x_{k+1} - \hat{x}_{k+1} ] \nonumber \\
    &= \bbe[ x_{k+1} ] - \bbe[ \hat{x}_{k+1} ] \nonumber \\
    &= \bbe \left[ A_k x_k + B_k K \hat{x}_k + w_k \right] - \bbe \left[ \Bar{A} \hat{x}_{k} + \Bar{B} K \hat{x}_k + L (y_k - \hat{y}_k) \right] \\
    &= \Bar{A} \bbe [x_k] + \Bar{B} K \bbe [\hat{x}_k] - (\Bar{A}  + \Bar{B} K) \bbe [\hat{x}_k] - L \bbe [ y_k - \hat{y}_k ] \nonumber \\
    &= \Bar{A} \bbe [x_k] + \Bar{B} K \bbe [\hat{x}_k] - (\Bar{A}  + \Bar{B} K) \bbe [\hat{x}_k] - L \Bar{C} \mathbb{E}[e_k] \nonumber \\
    &= \Bar{A} \bbe [x_k] - \Bar{A}  \bbe [\hat{x}_k] + \Bar{B} K \bbe [\hat{x}_k] - \Bar{B} K \bbe [\hat{x}_k] - L \Bar{C} \mathbb{E}[e_k] \nonumber \\
    &= (\Bar{A} - L \Bar{C}) \mathbb{E}[e_k] . \label{eq:first_moment_state_residual}
\end{align}

\subsection{Second moment dynamics}
For the state and state-estimate second moment dynamics, denote
\begin{alignat*}{2}
    & X_k = \vect \bbe \left[ x_{k} x_{k}^\top \right], \quad
    && \tilde{X}_k = \vect \bbe \left[ x_{k} \hat{x}_{k}^\top \right], \\
    & \breve{X}_k = \vect \bbe \left[ \hat{x}_{k} x_{k}^\top \right], \quad
    && \hat{X}_k = \vect \bbe \left[ \hat{x}_{k} \hat{x}_{k}^\top \right].
\end{alignat*}
We have
\begin{align}
    X_{k+1} 
    &=
    \vect \bbe \left[ x_{k+1} x_{k+1}^\top \right] \nonumber \\
    &=
    \vect \bbe \left[ (A_k x_k + B_k K \hat{x}_k + w_k) (A_k x_k + B_k K \hat{x}_k + w_k)^\top \right] \nonumber \\
    &=
    \left( \bar{A} \otimes \bar{A} + \Sigma_A^\prime \right) X_k + 
    ((\bar{B} K) \otimes \bar{A}) \tilde{X}_k
    + (\bar{A} \otimes (\bar{B} K)) \breve{X}_k + 
    \left( \bar{B} \otimes \bar{B} + \Sigma_B^\prime \right) (K \otimes K) \hat{X}_k +
    \vect(\Sigma_w), \label{eq:state_second_moment1}
\end{align}
and
\begin{align}
    \tilde{X}_{k+1}
    &=
    \vect \bbe \left[ x_{k+1} \hat{x}_{k+1}^\top \right] \nonumber \\
    &=
    \vect \bbe \left[ (A_k x_k + B_k K \hat{x}_k + w_k) \left( L C_k x_k + (\Bar{A} + \Bar{B} K - L \Bar{C} ) \hat{x}_{k} + L v_k \right)^\top \right] \nonumber \\
    &=
    ((L \Bar{C}) \otimes \Bar{A}) X_k 
    + \left( (\Bar{A} + \Bar{B} K - L \Bar{C}) \otimes \Bar{A} \right) \tilde{X}_k
    + ((L \Bar{C}) \otimes (\Bar{B} K) ) \breve{X}_k 
    + \left( (\Bar{A} + \Bar{B} K - L \Bar{C}) \otimes (\Bar{B} K) \right) \hat{X}_k, \label{eq:state_second_moment2}
\end{align}
and
\begin{align}
    \breve{X}_{k+1}
    &=
    \vect \bbe \left[ \hat{x}_{k+1} x_{k+1}^\top \right] \nonumber \\
    &=
    \vect \bbe \left[ \left( L C_k x_k + (\Bar{A} + \Bar{B} K - L \Bar{C} ) \hat{x}_{k} + L v_k \right) (A_k x_k + B_k K \hat{x}_k + w_k)^\top \right] \nonumber \\
    &=
    (\Bar{A} \otimes (L \Bar{C})) X_k 
    + ((\Bar{B} K) \otimes (L \Bar{C}) ) \tilde{X}_k 
    + \left( \Bar{A} \otimes (\Bar{A} + \Bar{B} K - L \Bar{C}) \right) \breve{X}_k
    + \left( (\Bar{B} K) \otimes (\Bar{A} + \Bar{B} K - L \Bar{C}) \right) \hat{X}_k, \label{eq:state_second_moment3}
\end{align}
and
\begin{align}
    \hat{X}_{k+1}
    &=
    \vect \bbe \left[ \hat{x}_{k+1} \hat{x}_{k+1}^\top \right] \nonumber \\
    &=
    \vect \bbe \left[ \left( L C_k x_k + (\Bar{A} + \Bar{B} K - L \Bar{C} ) \hat{x}_{k} + L v_k \right) \left( L C_k x_k + (\Bar{A} + \Bar{B} K - L \Bar{C} ) \hat{x}_{k} + L v_k \right)^\top \right] \nonumber \\
    &=
    (L \otimes L) (\bar{C} \otimes \bar{C} + \Sigma_C^\prime) X_k 
    + \left( (\Bar{A} + \Bar{B} K - L \Bar{C} ) \otimes (L \Bar{C}) \right) \tilde{X}_k \nonumber \\
    & \quad + \left( (L \Bar{C}) \otimes (\Bar{A} + \Bar{B} K - L \Bar{C} ) \right) \breve{X}_k
    + \left( (\Bar{A} + \Bar{B} K - L \Bar{C} ) \otimes (\Bar{A} + \Bar{B} K - L \Bar{C} ) \right) \hat{X}_k 
    + (L \otimes L) \vect(\Sigma_v). \label{eq:state_second_moment4}
\end{align}
Define
\begin{align*}
    \mathcal{X}_k \Let 
    \begin{bmatrix}
    X_k \\
    \tilde{X}_k \\
    \breve{X}_k \\
    \hat{X}_k
    \end{bmatrix}, \qquad \text{and} \qquad
    \mathcal{V}
    \Let
    \begin{bmatrix}
    \vect (\Sigma_w) \\
    \vect (\Sigma_v)
    \end{bmatrix}.
\end{align*}
By gathering the matrix coefficients in equations \eqref{eq:state_second_moment1}, \eqref{eq:state_second_moment2}, \eqref{eq:state_second_moment3}, \eqref{eq:state_second_moment4} as
\begin{align*}
    H \Let 
    \begin{bmatrix}
    \bar{A} \otimes \bar{A} + \Sigma_A^\prime \! & \!(\bar{B} K) \otimes \bar{A}\! & \!\bar{A} \otimes (\bar{B} K)\! & \!\left( \bar{B} \otimes \bar{B} + \Sigma_B^\prime \right) (K \otimes K) \\
    (L \Bar{C}) \otimes \Bar{A}\! & \!(\Bar{A} + \Bar{B} K - L \Bar{C}) \otimes \Bar{A}\! & \!(L \Bar{C}) \otimes (\Bar{B} K)\! & \!(\Bar{A} + \Bar{B} K - L \Bar{C}) \otimes (\Bar{B} K) \\
    \Bar{A} \otimes (L \Bar{C})\! & \!(\Bar{B} K) \otimes (L \Bar{C})\! & \!\Bar{A} \otimes (\Bar{A} + \Bar{B} K - L \Bar{C})\! & \!(\Bar{B} K) \otimes (\Bar{A} + \Bar{B} K - L \Bar{C}) \\
    (L \otimes L) (\bar{C} \otimes \bar{C} + \Sigma_C^\prime)\! & \!(\Bar{A} + \Bar{B} K - L \Bar{C} ) \otimes (L \Bar{C})\! & \!(L \Bar{C}) \otimes (\Bar{A} + \Bar{B} K - L \Bar{C} )\! & \!(\Bar{A} + \Bar{B} K - L \Bar{C} ) \otimes (\Bar{A} + \Bar{B} K - L \Bar{C} )
    \end{bmatrix}
\end{align*}
and
\begin{align*}
    \Phi 
    \Let
    \begin{bmatrix}
    I_{n} \otimes I_{n} & 0_{n^{2} \times 1} \\
    0_{n^{2} \times n^{2}} & 0_{n^{2} \times 1} \\
    0_{n^{2} \times n^{2}} & 0_{n^{2} \times 1} \\
    0_{n^{2} \times n^{2}} & L \otimes L
    \end{bmatrix}
\end{align*}
we have the compact representation of \eqref{eq:state_second_moment1}, \eqref{eq:state_second_moment2}, \eqref{eq:state_second_moment3}, \eqref{eq:state_second_moment4} as
\begin{align}
    \mathcal{X}_{k+1} = H \mathcal{X}_k + \Phi \mathcal{V}. \label{eq:state_second_moment_compact}
\end{align}
For the state- and output-residual second moments, denote $E_k = \vect \bbe \left[ e_{k} e_{k}^\top \right]
R_k = \vect \bbe \left[ r_{k} r_{k}^\top \right]$. We have
\begin{align}
    E_{k}
    &=
    \vect \bbe \left[ e_{k} e_{k}^\top \right] \nonumber \\
    &=
    \vect \bbe \left[ (x_k - \hat{x}_k) (x_k - \hat{x}_k)^\top \right] \nonumber \\
    &=
    X_k - \tilde{X}_k - \breve{X}_k + \hat{X}_k \label{eq:state_residual_cov}
\end{align}
and
\begin{align}
    R_{k}
    &=
    \vect \bbe \left[ r_{k} r_{k}^\top \right] \nonumber \\
    &=
    \vect \bbe \left[ (y_k - \hat{y}_k) (y_k - \hat{y}_k)^\top \right] \nonumber \\    
    &=
    \vect \bbe \left[ (C_k x_k + v_k - \Bar{C} \hat{x}_k) (C_k x_k + v_k - \Bar{C} \hat{x}_k)^\top \right] \nonumber \\ 
    &=
    \vect \bbe \left[ (\Bar{C} (x_k - \hat{x}_k) + (C_k - \Bar{C} ) x_k + v_k ) (\Bar{C} (x_k - \hat{x}_k) + (C_k - \Bar{C} ) x_k + v_k)^\top \right] \nonumber \\
    &=
    (\Bar{C} \otimes \Bar{C}) E_k + \Sigma_C^\prime X_k + \vect(\Sigma_v) \label{eq:output_residual_cov}
\end{align}

\end{document}